Sequence analysis

# A statistical physics perspective on alignment-independent protein sequence comparison

Amit K. Chattopadhyay[1], Diar Nasiev[1] and Darren R. Flower[2,*]

[1]School of Engineering and Applied Science, Nonlinearity and Complexity Research Group and [2]School of Life and Health Sciences, University of Aston, Aston Triangle, Birmingham, UK

*To whom correspondence should be addressed.
Associate Editor: John Hancock

## Abstract

**Motivation**: Within bioinformatics, the textual alignment of amino acid sequences has long dominated the determination of similarity between proteins, with all that implies for shared structure, function and evolutionary descent. Despite the relative success of modern-day sequence alignment algorithms, so-called alignment-free approaches offer a complementary means of determining and expressing similarity, with potential benefits in certain key applications, such as regression analysis of protein structure-function studies, where alignment-base similarity has performed poorly.
**Results**: Here, we offer a fresh, statistical physics-based perspective focusing on the question of alignment-free comparison, in the process adapting results from 'first passage probability distribution' to summarize statistics of ensemble averaged amino acid propensity values. In this article, we introduce and elaborate this approach.
**Contact**: d.r.flower@aston.ac.uk

## 1 Introduction

Determining the similarity between macromolecules is central to bioinformatics. While comparison of 3-dimensional macromolecular structures remains an active area, most work focuses on macromolecular sequences. From sequence similarity devolves much of our understanding of evolutionary homology and probable structural and functional relatedness, allowing sequences to be grouped in a meaningful way. It is the basis of inherited or inferred functional annotation, allowing us to deduce the broad function of proteins in newly sequenced genomes.

Similarity is determined, almost exclusively, through the alignment of sequences as text. Textual sequence similarity is taken as a surrogate for common ancestry and, by extension, functional and structural similarity. Most approaches to protein sequence similarity use models of sequence evolution and compare amino-acid strings, searching for linear conservation of sequence.

Sequence alignment, where equivalent or near-equivalent symbols are brought into register, has been investigated intensely for many decades (Altschul, 1991; Vinga and Almeida, 2003), and thus an enormous associated literature has accumulated. Typically, substitution matrices specify a score for aligning pairs of nucleotides or amino acids; in such matrices different amino acids or nucleotides score differently according to the potential likelihood that one will replace the other in a sequence. For amino acids, many matrices have been published, based on many rationales (Feng et al., 1984; Taylor, 1986), including the genetic code and amino acid physicochemical properties. Most commonly-used matrices are typically derived empirically from exhaustive comparison of known sequences or structures.

The log-odds matrices (Schwartz and Dayhoff, 1978) derived from the PAM model of protein evolution (Dayhoff et al., 1978) was, for many years, the most widely used. Statistical results indicate such matrices adopt an implicit 'log-odds' form, with a specific target distribution for aligned residue pairs. The sensitivity of protein sequence searching depends on the selection of appropriate substitution matrices (Henikoff and Henikoff, 1992; 1993; Pearson, 1995). BLOSUM, and other commonly-used matrices, constructed from particular sets of related proteins, are tailored to target frequencies reflecting implied standard background amino acid compositions.

While probabilistic extensions to sequence alignment, such as profiles and Hidden Markov Models (HMMs), can capture position-specific variation in multiple alignments, and typically demonstrate significantly augmented sensitivity, all alignment methods remain prone to similar limitations.

Compared to sequence alignment, alternative approaches, grouped together as alignment-free techniques, and first proposed by Blaisdell (1986), have not been investigated as thoroughly (Davies et al., 2007). Extant methods fall into several groups; of these, perhaps the most explored approach focuses on sequence comparison based on the joint sub-word or k-tuple content of groups of sequences, and their analysis using increasingly sophisticated probabilistic statistics.

Amongst other approaches, methods based on plotting so-called propensity scales (Nakai et al., 1988) have enjoyed long-standing popularity; with scales mirroring one or more amino acid properties, such as hydrophobicity (Hopp and Woods, 1981) or electronegativity. Such scales abound: AAindex has collected 545 different published scales (Kawashima et al., 2008). Others have summarized such data, producing, inter alia, three (Hellberg et al., 1987) or five scales (Sandberg et al., 1998; Venkatarajan and Braun, 2001).

A propensity scale is a means to characterize numerically local sequence properties, usually plotting amino acid properties along a sequence. Plotting individual values is of little value however unless there is an obvious periodicity. More helpful is to average the values using a moving or sliding window: typically using a flat, symmetrical window of no more than 10–20 amino acids in length. A moving window can have several potential parameters: first, the scale chosen; second, the window length; third, the window shape; fourth, whether the window is symmetrical or unbalanced; and fifthly, how values are smoothed after averaging. Typically, windows are short, flat, and symmetrical about the central residue, which takes the averaged value. Several different window shapes have been suggested, corresponding to weighting each position independently.

Independent of the exact parameters used to define the window function, various kinds of smoothing are available, including digital filtering, integral transforms (Fourier and Cosine), and wavelet analysis, each with their own characteristics. Smoothing seeks to reduce the random component of the initial value spectrum generated by windowing, with the high-frequency regions removed, leaving only dominant low-frequency modes.

The value of individual propensity plots is limited. It works well, say, for predicting transmembrane regions within proteins, where peaks in the plot can correlate well with regions highly enriched in hydrophobic residues. Hitherto, it has proved difficult to interpret such plots other than by a peak-spotting. Several decades ago, most predictors were based on identifying maximally valued regions of sequences; essentially looking for peaks, or troughs, in some form of a propensity plot. Epitopes, such as immunological T- or B-cell epitopes (Deavin et al., 1996; Hopp and Woods, 1981); loops and surface exposure (Dovidchenko et al., 2008); and transmembrane helices (Sipos and von Heijne, 1993) were—and often still are—predicted this way. Propensity scales have also been used in QSAR studies, particularly those focusing on peptides (Hellberg et al., 1987; Sandberg et al., 1998).

In what follows, we take a decisive step away from such analyses, using techniques drawn from statistical physics. Specifically, we use techniques from the rich literature of first passage probability distribution, sometimes referred to as the 'persistence' problem (Bush and Chattopadhyay, 2014; Derrida et al., 1995; Majumdar, 1996; Bray, 2013; Redner, 2007); applying them for the first time to sequence analysis. Persistence analysis has found applications in many fields including, inter alia: stock market analysis (Ren, 2005), modeling immunological systems (Chattopadhyay and Burroughs, 2007), extremal value statistics pertaining to data degradation (Whitmore, 1986), modeling the population biology of HIV dynamics (Tuckwell and Wan, 2000), event detection time for mobile sensor networks (Inaltekin et al., 2007), and supply chain optimization (Wakuta, 2000).

In this article, we develop and analyze propensity data by modeling sequences as a time-series, estimating the scaling regime of a generalized probability density function (PDF) of a variable derived from the original propensity data structure. Our protocol enables us to abstract key features from propensity plots while remaining free of any text-based alignment scheme. We then apply this alignment-independent approach to the analysis of protein sequences, evaluating it as a potential means of automatically characterizing and clustering large numbers of sequences.

## 2 Persistence Analysis: Deriving Order Parameters from Protein Sequences

The idea underlying persistence analysis is simple. It relies on the nature of the PDF of stochastic time series data $X(t)$: in our case, ensemble-averaged propensity scale data. The basic question asked is: what is the probability $p(t)$ that the field $X(t)$ has not changed sign up to time $t$, starting from an initial configuration $X(t_0)$? An equivalent question would be: what is the probability $p(t_1,t_2)$ that the field $X(t)$ changes sign N times between $t_1$ and $t_2$ for $X(t) > <X>$? In our case, $<X> = X_0$ or the mean value of the time series data $X(t)$. Depending on the nature of the 'random walk'—in this case a protein sequence— this probability is given by $p(t_1,t_2) \sim (t_1/t_2)^{-m}$; where the exponent 'm' assumes different values depending on whether this probability is calculated using data above or below a certain threshold. The threshold is typically the mean value ($X_0$) of the stochastic data. Thus multiple 'order parameters' may emerge from the same description depending on the chosen value of 'm'.

Our approach utilizes 'extremal value statistics' to analyze sequence structure. This builds on a tacit first approximation that sequences—as represented by propensity plots—can be seen as being predominantly stochastic in structure, at least when viewed synoptically. At this stage, the focus is on a specific manifestation, that of the 'first passage probability' distribution around the mean value $X_0$ of the data points $X(t)$ where 't' is our metaphor for the location of the data point in the time series-like sequence. What this 'first passage statistics' captures is the distribution of the 'return times' of the fluctuating 'time series' across the given threshold $X(t) = X_0$.

Figure 1 shows the 'return time lengths' from data points below the line $X(t) = X_0$ to one above this line and then back again to $X(t) < X_0$. As an example, $t_+^{(1)}$ defines the first 'plus'-type return time where 'plus' refers to the regime $X(t) > X_0$ such that the time series starts from a point below this line $(X(t) < X_0)$ and after crossing this line returns back to the regime $X(t) < X_0$. Similarly, $t_+^{(2)}$ refers to the second such 'plus-type' return time; and so on.

$t_-^{(1)}$ refers instead to the first 'minus-type' return time where the time series count starts at $X(t) > X_0$, then crosses the threshold line in to the $X(t) < X_0$ regime and finally returns to $X(t) > X_0$.

Our interest is in the probability distributions of the $t_+$ and $t_-$ transitions. Such a probability distribution is achieved by calculating histograms of 'return times' above ($t_+$) and below ($t_-$) the line $X(t) = X_0$. The histograms are then normalized to obtain the desired PDFs. We assume that sequences are, to a first approximation, 'inherently random'. As long as we are within the stochastic realm, all prior statistical results, including that for long range correlated stochastic data, suggest the existence of universal power law exponents defining the first passage probability distribution statistics related to $t_+$ and $t_-$ transitions. In our description, we will use $m_+$ to identify the exponent for $X(t) > X_0$ and $m_-$ for the exponent characterizing $X(t) < X_0$. We now focus on this two dimensional $(m_n, i)$ plane where $n = +, -$.

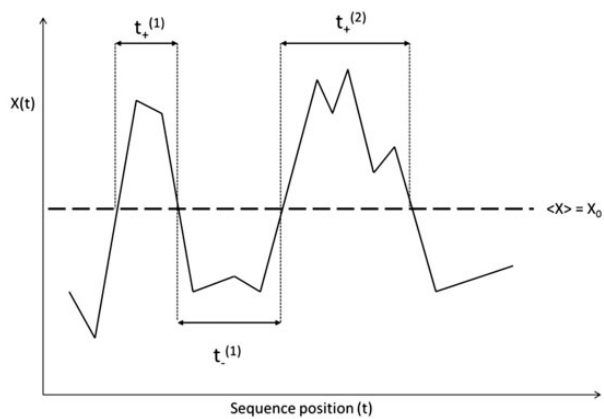

**Fig. 1.** An illustration of first passage probability distribution across a threshold $X(t) = X_0$

## 3 Methods

Here, we characterize protein sequences using an alignment-free approach based on techniques of time series analysis commonly-used in statistical physics and elsewhere (Bray, 2013; Redner, 2007).

### 3.1 Plotting of propensity data
In general, a window could adopt any arbitrary shape; thus we may assume a generalized window will have this form:

$$\overline{V}_i^p = w_k S_i^p + \sum_{k=1}^{n} w_k S_{i-k}^p + \sum_{l=1}^{m} w_l S_{i+l}^p$$

Within the window there will be $m$ amino acids that are to be averaged upstream of the target residue, and will have $n$ amino acids downstream, plus the value for the residue itself; all residues within the window will have independent and arbitrary coefficients ($w$) that will selectively weight individual positions within the travelling window. Averaging can be undertaken several times: values averaged over any particular iteration being the values generated by averaging in the previous iteration.

We extracted 544 useable propensity values from AAIndex (Kawashima *et al.*, 2008). All scales were used. Sequences were converted to numerical profiles comprising propensity values. Using a window length of 7, one round of uniform smoothing was used per sequence. Each sequence behaved as intrinsically stochastic time series-like data. For each sequence processed, we generated 544 different smoothed profiles, corresponding to the 544 scales from AAindex. Each of these 544 datasets was arranged in $N$ column vectors, where each column vector represented a sequence. The number of entries in each of these column vectors was identical. The block structure is a matrix $M$ ($N \times 544$) where each element of this matrix is a column vector.

### 3.2 Propensity analysis
For each sequence drawn from Pfam (Finn *et al.*, 2014), and for each of the 544 AAindex propensity scales, separate $m_+$ and $m_-$ values were calculated from the corresponding $t_+$ and $t_-$ data. For each sequence, the arithmetic means of $544\, m_+$ and $544\, m_-$ values were derived: this gave the 'order parameter' $m = 0.5*(m_+ + m_-)$. This is akin to statistical 'ensemble averaging', including any cogent non-ergodicity of the ensemble. The order parameter $m$ thus represents each protein sequence as a single number. The resulting $m$-values were then used to cluster the sequences using the scheme detailed below. The schematic algorithm for calculating the values $m_+$ and $m_-$ value for each of $N$ sequences is shown below:

1. $(m_+, m_-)$ calculated for each of the 544 column vectors of row $i$ of the $M$-matrix
2. Ensemble average (arithmetical mean) taken of $(m_+, m_-)$ for all 544 vectors in row $i$ of the $M$-matrix
3. Steps 1–2 repeated $N$ times, for all 544 vectors in each row, to generate $N$-sets of $(m_+, m_-)$

Our 'order parameter' is thus the mean of the previously defined $m_+$ and $m_-$ values obtained individually as 'scaling exponents' ('persistence exponents') of the $t_+$ (or $t_-$) PDFs. The ultimate objective of this analysis was a clear clustering of the protein sequences such that similar proteins fall into separate clusters that are defined by separate combinations of $(m_n, i)$ values (coordinate location in the $m_n$–$i$ space) where $n = +, -$ and $i$ = position of the protein sequence along the $x$-axis. Clustering follows the logic below:

1. Plot $(m_+, m_-)$ versus relevant sequence number to generate the phase diagram
2. Grouping of $(m_+, m_-)$ versus sequence number $i$ in the phase diagram
3. Starting from the phase diagram defined in 2, estimate $dm = mi + 1 - mi$, $f = (1 + dm)*di$ between every two points representing each sequence.
4. Plot $f$ versus $i$ using a first adjustable threshold (mean of the separation distances in the phase diagram plot) that separates out the large $f$'s form the small $f$'s. Use dendogram based MATLAB clustering protocol
5. Since the dendogram generates more clusters than the system actually has, use a second threshold (standard deviation of the data points in the phase diagram plot) and repeat step 2. The clustering accuracy will be shown by the level of gradient equality (see Fig. 3)

Clustering data are then plotted sequentially to evaluate the size of successive clusters. The PDFs of cluster lengths exhibit power-law scaling. More importantly, the scaling exponents can be used to compare the clustering accuracy of the Matlab-inbuilt architecture. Comparing this to known clusters provides the probabilistic values of the two external parameters determining the eventual cluster quality.

### 3.3 Application to test cases
We extracted one arbitrary but representative sequence family from each superfamily in the Pfam database (version 27.0, March 2013, 14831 families; Finn *et al.*, 2014). Seed sequences were used in preference to final sequence sets, for reasons of reliability, since members were chosen by human experts without the involvement of potentially-unreliable automated methods. Downloaded alignments were converted to un-gapped sequence sets.

Figure 3 is an accuracy-check of our algorithm, estimating the cluster size distribution D. The gradients of the dotted and dashed lines in the log-log plots (cluster PDF of size) in Figure 3 are compared with that of the solid line (actual data) in identifying the 'best

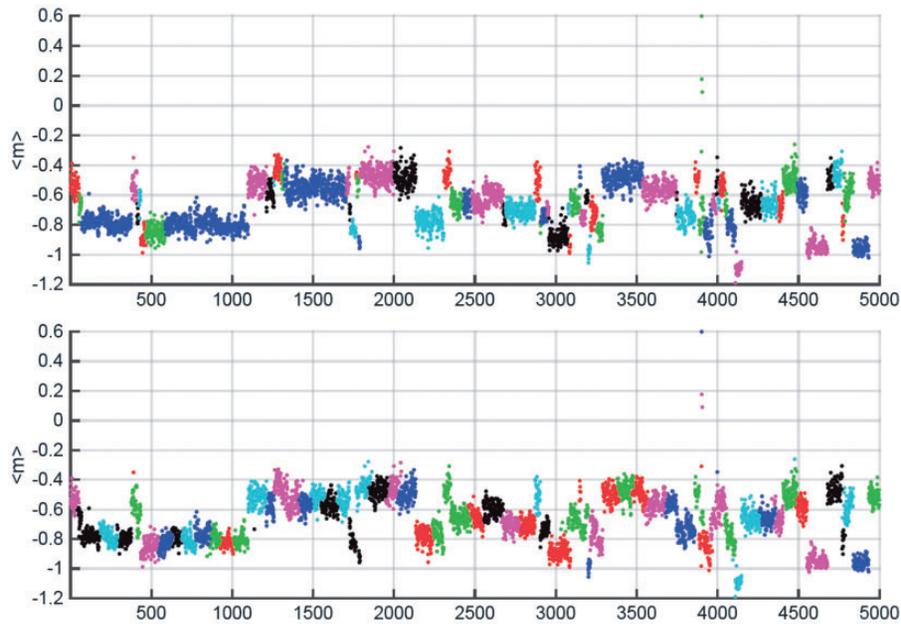

**Fig. 2.** Clustering from the part of the Pfam dataset. Panel 1 represents the reference set of known clusters with groupings obtained from Pfam. Panel 2 is the clustering obtained using an inbuilt Matlab algorithm which uses only the number of clusters from Pfam (Panel 1) as input. The y-axis represents the average value of the order parameter m, while the x-axis represents the sequence. Separate clusters are shown using different colours

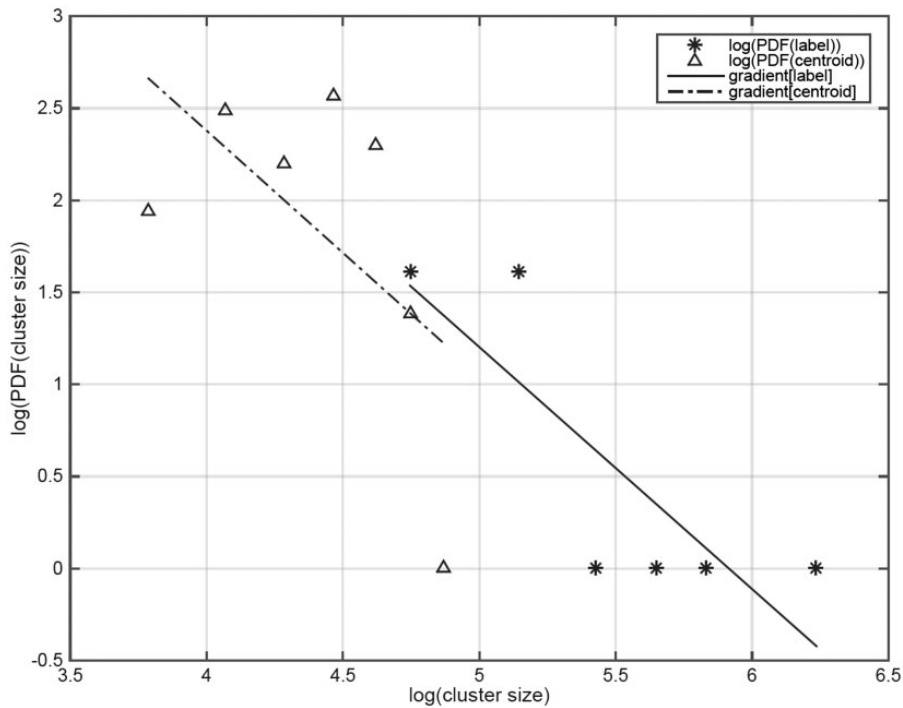

**Fig. 3.** Histograms of cluster sizes obtained from the clustering data in Figure 2 and plotted on a log-log scale. The solid line represents actual data (panel 1 of Fig. 2), the dotted line represents the result from the Matlab-inbuilt algorithm (panel 2 of Fig. 2)

fit' (least squares) structure that leads to the most optimized clustering scheme, as shown in Figure 2.

The distribution follows a power-law profile ($D \sim S^{-p}$) with an (super-diffusive) exponent $p \sim 1$. The two (threshold) parameters are tuned to predict the optimal fit straight line from our algorithm parallel to that of the 'known' data from panel 1. To the best of our knowledge, no extant procedure has such high accuracy with only two tunable parameters.

The Matlab-based dendogram algorithm used requires the number of clusters as input, but beyond that the resulting clustering, including the size and location of clusters, are independent and no assumptions are made or prior knowledge used. As expected, the location and size of resulting clusters is not 100% accurate, yet it is clear from Figure 2 that mismatches are generally marginal. We also calculated two widely-used quality indices as summary measures of the Pfam clustering. The cluster separation measure of Davies and Bouldin (1979) gave a value of 0.75032, while the Dunn compactness index was 0.03296. Both values are consistent with excellent overall clustering.

The data in Figures 2 and 3 establish the robustness of our algorithm. Using sequence-data with different overall characteristics and number of independent clusters (shown in panel 1 of Fig. 2: our benchmark), our results show that irrespective of the low accuracy of the cluster distribution gradient value ('p'), as clearly seen from the non-parallel lines of Figure 3, cluster identification, results in highly accurate discrimination of groups. In other work, concentrating on more readily-apparent similarity within single Pfam clans, we have found using our current minimal and unoptimized method similar reliability at a much finer granularity. Together, these results demonstrate this sequence representation is both consistent and discriminatory at both high and low granularity.

## 4 Discussion

The analysis of sequence similarity is the cornerstone upon which much of bioinformatics is built. Hitherto, alignment-based approaches have completely dominated work in this area, while other approaches, of which there are several (Davies *et al.*, 2007), have been examined with much less thoroughness. Yet the need for effective approaches, able to transcend the limitations of text matching, is clear, if not widely appreciated.

The protocol described here takes propensity plots produced from the 544 scales in AAindex (Kawashima *et al.*, 2008), averages them, and generates a single value characteristic of a whole sequence. Our implementation is, in essence, an out-of-the-box application of existing results, with immense potential for future refinement. We have applied here a well-understood statistical method from stochastic mathematics, using it to interpret and identify independent clusters in protein sequence data. Our approach introduces a fundamentally new way to represent sequences which is nonetheless founded on the long-standing concept of propensity scales, and, capitalizing on features of this representation, we have used it to power a novel approach to clustering. This method is able to capture much of the overall structure of a propensity plot in a single but discriminatory and self-consistent value.

Many properties of a protein are encoded within its sequence in a subtle and recondite manner not amendable to direct identification through sequence alignment or the recognition of characteristic sequence motifs. Likewise, the discovery of functionally-similar but sequence-distinct proteins may be frustrated by a lack of ostensive similarity to proteins of known provenance. In such a situation, alignment-dependent approaches may produce ambiguous results or fail completely.

There are many examples where structural or functional similarities are readily apparent experimentally, yet are difficult—if not impossible—to detect from textual sequence alignment. Perhaps the most obvious are the so-called twilight and midnight zones (Rost, 1999). Most protein sequences will fold into unique 3-dimensional structures, and similar sequences will typically have similar structures. Sequence alignments can routinely distinguish between sequence pairs known to have similar or non-similar structures when the sequence identity is greater than >40%. This unequivocal signal becomes lost at 20–35% sequence identity: the so-called twilight zone. Alignment-methods often fail to align paired amino-acids correctly even with 20–30% identity. Structural alignment has revealed many, many examples of so-called structural superfamilies, where proteins with less than 10% identity nonetheless retain structural propinquity (Flower, 1993; Flower *et al.*, 1993). The average identity between all sequence pairs of related structures is 8–10%, and this marks the midnight zone, which is predominantly populated by protein structures that have become similar by convergent evolution.

Thus, we need to apply our approach to a variety of both solved and unsolved problems to explore its value. Solved problems include searching within and beyond the twilight zone, and to classify and identify structural and functional relationships within it effectively, it is imperative to explore alternative approaches to pairwise similarity and BLAST statistics (Karlin and Altschul, 1990). Unsolved problems include clustering whole genomes where the result is unknown, and is complicated by the presence of multi domain proteins, internal repeats, *etc.*; the development of regression models using our order parameter values as descriptors; and a surrogate of sequence searching using order values to define similarity measures.

In protein design there is a need to move beyond making piecemeal changes to extant sequences to identify wholly new sequences with new functions and structures. Effective alignment free approaches should allow us to address such issues, particularly through the development of properly grounded regression approaches to protein sequence analysis. There are many examples of such approaches: the proteochemometric analysis of protein sequences (van Westen *et al.*, 2013), the prediction of candidate vaccines (Doytchinova and Flower, 2007a,b), and the successful assignment of bacterial proteins to various subcellular locations (Sjöström *et al.*, 1995).

Regression approaches typically require three viable components: an induction engine (based on multivariate statistics, such as PLS, or a machine learning algorithm), data to be modeled (which can be quantitative or categorical in nature), and an appropriate data representation. Data modeling methods have reached sufficient maturity, and data quality is constrained by its availability on a case-by-case basis, so increasingly it is the choice of data representation that is the crucial arbiter of success. This is especially true for protein regression, which lags far behind equivalent work for small molecules. Alignment-independent similarity measures, such as our representation, offer an interesting and seemingly productive avenue for achieving progress in this endeavor.

To go beyond sequence motifs and profiles, HMMs and like methodology—and thus identify common function, structure, and evolution—new, distinct, yet complimentary, methods must be devised: alignment-free methods that can work with textual alignment to identity similarity manifest as shared structure and function.

In this article, we have used results from statistical physics to address alignment-free comparison, adapting results from 'first passage probability distribution' to derive a single summary value able to differentiate sequence groups with high accuracy at several levels of granularity. This approach is potential highly robust being largely independent of fluctuation in the tunable parameters. We anticipate that this approach will ultimately take its place alongside textual alignment as a strongly complimentary method for sequence analysis, with many advantages compared to conventional techniques.

## Acknowledgement

We thank Professor T. K. Attwood for helpful discussions.

## Funding

This work was supported by Aston University through internal departmental funding.

*Conflict of Interest*: none declared.

## References

Altschul,S.F. (1991) Amino acid substitution matrices from an information theoretic perspective. *J. Mol. Biol.*, 219, 555–565.

Blaisdell,B.E. (1986) A Measure of the similarity of sets of sequences not requiring sequence alignment. *Proc. Natl. Acad. Sci. USA*, 83, 5155–5159.

Bray,A.J. *et al.* (2013) Persistence and first passage properties in non-equilibrium systems. *Adv. Phys.*, 62, 225–361.

Bush,D.J. and Chattopadhyay,A.K. (2014) Contact time periods in immunological synapse. *Physical Review E*, 90, 042706.

Chattopadhyay,A.K. and Burroughs,N. (2007) Close contact fluctuations: the seeding of signaling domains in immunological synapse. *Europhys. Lett.*, 77, 48003.

Davies,D.L. and Bouldin,D.W. (1979) A cluster separation measure. *IEEE Trans. Pattern Anal. Mach. Intell.*, 1, 224–227.

Davies,M.N. *et al.* (2007) Proteomic applications of automated GPCR classification. *Proteomics*, 7, 2800–2814.

Dayhoff,M.O. *et al.* (1978) A model of Evolutionary change in proteins. In: Dayhoff,M.O. (ed.) Atlas of Protein sequence and structure. National Biomedical Research Foundation, pp. 345–352.

Deavin,A.J. *et al.* (1996) Statistical comparison of established T-cell epitope predictors against a large database of human and murine antigens. *Mol. Immunol.*, 33, 145–155.

Derrida,B. *et al.* (1995) Exact first-passage exponents of 1D domain growth: relation to a reaction-diffusion model. *Phys. Rev. Lett.*, 75, 751–754.

Dovidchenko,N.V. *et al.* (2008) Prediction of loop regions in protein sequence. *J. Bioinform. Comput. Biol.*, 6, 1035–1047.

Doytchinova,I.A. and Flower,D.R. (2007a) VaxiJen: a server for prediction of protective antigens, tumour antigens and subunit vaccines. *BMC Bioinformatics*, 8, 4.

Doytchinova,I.A. and Flower,D.R. (2007b) Identifying candidate subunit vaccines using an alignment-independent method based on principal amino acid properties. *Vaccine*, 25, 856–866.

Dunn,J.C. (1973) A fuzzy relative of the ISODATA process and its use in detecting compact well-separated clusters. *J. Cybern.*, 3, 32–57.

Feng,D.F. *et al.* (1984) Aligning amino acid sequences: comparison of commonly used methods. *J. Mol. Evol.*, 21, 112–125.

Finn,R.D. *et al.* (2014) Pfam: the protein families database. *Nucleic Acids Res.*, 42, D222–D2230.

Flower,D.R. *et al.* (1993) Structure and sequence relationships in the lipocalins and related proteins. *Protein Sci.*, 2, 753–761.

Flower,D.R. (1993) Structural relationship of streptavidin to the calycin protein superfamily. *FEBS Lett.*, 333, 99–102.

Hellberg,S. *et al.* (1987) Peptide quantitative structure-activity relationships, a multivariate approach. *J. Med. Chem.*, 30, 1126–1135.

Henikoff,S. and Henikoff,J.G. (1992) Amino acid substitution matrices from protein blocks. *Proc. Natl. Acad. Sci. USA* 89, 10915–10919.

Henikoff,S. and Henikoff,J.G. (1993) Performance evaluation of amino acid substitution matrices. *Proteins*, 17, 49–61.

Hopp,T.P. and Woods,K.R. (1981) Prediction of protein antigenic determinants from amino acid sequences. *Proc. Natl. Acad. Sci. USA*, 78, 3824–3828.

Inaltekin,H. *et al.* (2007) Event detection time for mobile sensor networks using first passage processes. *IEEE Global Telecom. Conf.*, 1174–1179.

Karlin,S. and Altschul,S.F. (1990) Methods for assessing the statistical significance of molecular sequence features by using general scoring schemes. *Proc. Natl. Acad. Sci. USA*, 87, 2264–2268.

Kawashima,S. *et al.* (2008) AAindex: amino acid index database, progress report 2008. *Nucleic Acids Res.*, 36, D202–D205.

Majumdar,S.N. *et al.* (1996) Global persistence exponent for nonequilibrium critical dynamics. *Phys. Rev. Lett.*, 77, 3704–3707.

Nakai,K. *et al.* (1988) Cluster analysis of amino acid indices for prediction of protein structure and function. *Protein Eng.*, 2, 93–100.

Pearson,W.R. (1995) Comparison of methods for searching protein sequence databases. *Protein Sci.*, 4, 1145–1160.

Redner,S. (2007) A guide to first passage processes. Cambridge University Press ISBN: 978-0521036917.

Ren,F. *et al.* (2005) Persistence probabilities of the German DAX and Shanghai Index. *Physica A*, 350, 439–450.

Rost,B. (1999) Twilight zone of protein sequence alignments. *Protein Eng.*, 12, 85–94.

Sandberg,M. *et al.* (1998) New chemical descriptors relevant for the design of biologically active peptides. A multivariate characterization of 87 amino acids. *J. Med. Chem.*, 41, 2481–2491.

Schwartz,R.M. and Dayhoff,M.O. (1978) In: Dayhoff,M.O. (ed.) Matrices for detecting distant relationships in Atlas of Protein Sequence and Structure. pp. 353–358.

Sipos,L. and von Heijne,G. (1993) Predicting the topology of eukaryotic membrane proteins. *Eur. J. Biochem.*, 213, 1333–1340.

Sjöström,M. *et al.* (1995) Polypeptide sequence property relationships in *Escherichia coli* based on auto cross covariances. *Chemometr. Intell. Lab. Syst.*, 29, 295–305.

Taylor,W.R. (1986) The classification of amino acid conservation. *J. Theor. Biol.*, 119, 205–218.

Tuckwell,H.C. and Wan,F.Y.M. (2000) First passage time to detection in stochastic population dynamical models for HIV-1. *Appl. Math. Lett.*, 13, 79–83.

van Westen,G.J. *et al.* (2013) Benchmarking of protein descriptor sets in proteochemometric modeling (part 2): modeling performance of 13 amino acid descriptor sets. *J. Cheminform.*, 5, 42.

Venkatarajan,M.S. and Braun,W. (2001) New quantitative descriptors of amino acids based on multidimensional scaling of a large number of physical-chemical properties. *J. Mol. Model.*, 7, 445–453.

Vinga,S. and Almeida,J. (2003) Alignment-free sequence comparison-a review. *Bioinformatics*, 19, 513–523.

Wakuta,K. (2000) A first passage problem with multiple costs. *Math. Models Oper. Res.*, 51, 419–432.

Whitmore,G.A. (1986) First passage time models for duration data regression structures and competing risks. *Statistician*, 35, 207–219.